\documentclass[reprint, onecolumn,12pt]{revtex4-2}
\usepackage{amsmath,amssymb}
\usepackage{graphicx}
\usepackage{bm}
\usepackage{color}

\begin{document}

\preprint{APS/123-QED}


\title{Electronic cloaking of confined states in phosphorene junctions}

\author{S. Molina-Valdovinos}
\email{sergiomv@uaz.edu.mx}
\affiliation{Unidad Acad\'emica de Ciencia y Tecnolog\'ia de la Luz y la Materia, Universidad Aut\'onoma de Zacatecas, Carretera Zacatecas-Guadalajara Km. 6, Ejido La Escondida, 98160 Zacatecas, Zacatecas, M\'exico.}

\author{K. J. Lamas-Mart\'inez}
\affiliation{Unidad Acad\'emica de Ciencia y Tecnolog\'ia de la Luz y la Materia, Universidad Aut\'onoma de Zacatecas, Carretera Zacatecas-Guadalajara Km. 6, Ejido La Escondida, 98160 Zacatecas, Zacatecas, M\'exico.}

\author{J. A. Briones-Torres}%
\affiliation{Unidad Acad\'emica de Ciencia y Tecnolog\'ia de la Luz y la Materia, Universidad Aut\'onoma de Zacatecas, Carretera Zacatecas-Guadalajara Km. 6, Ejido La Escondida, 98160 Zacatecas, Zacatecas, M\'exico.}

\author{I. Rodr\'iguez-Vargas}%
\email{isaac@uaz.edu.mx}
\affiliation{Unidad Acad\'emica de Ciencia y Tecnolog\'ia de la Luz y la Materia, Universidad Aut\'onoma de Zacatecas, Carretera Zacatecas-Guadalajara Km. 6, Ejido La Escondida, 98160 Zacatecas, Zacatecas, M\'exico.}

\date{\today}

\begin{abstract}
We study the electronic transport of armchair and zigzag gated phosphorene junctions. We find confined states for both direction-dependent phosphorene junctions. In the case of armchair junctions confined states are reflected in the transmission properties as Fabry-P\'erot resonances at normal and oblique incidence. In the case of zigzag junctions confined states are invisible at normal incidence, resulting in a null transmission. At oblique incidence Fabry-P\'erot resonances are presented in the transmission as in the case of armchair junctions. This invisibility or electronic cloaking is related to the highly direction-dependent pseudospin texture of the charge carriers in phosphorene. Electronic cloaking is also manifested as a series of singular peaks in the conductance and as inverted peaks in the Seebeck coefficient. The characteristics of electronic cloaking are also susceptible to the modulation of the phosphorene bandgap and an external magnetic field. So, electronic cloaking in phosphorene junctions in principle could be tested through transport, thermoelectric or magnetotransport measurements. 
\end{abstract}


\keywords{Phosphorene, cloaked states, bounded states}

\maketitle



Cloaking effect consists in making objects invisible to radiation, acoustic waves, matter waves, heat and charge fluxes, among others~\cite{PhysRevApplied.4.037001,Raza_2016,PhysRevLett.113.205501,PhysRevLett.100.123002}. The typical cloaking effect is based on guiding plane waves around an object. The effect also refers to hiding an object in space. Cloaking effect is a phenomenon that can have a plethora of applications such as mantle cloaking, antennas, invisible sensors that do not perturb the field that they measure, etc.~\cite{PhysRevApplied.4.037001}. With the arrival of 2D materials, in particular graphene, there were reports of the cloaking effect in bilayer graphene junctions~\cite{NanGu-PhysRevLett.107.156603,Daboussi-pssb.201600430,MAITI2018330} and graphene nanoribbons~\cite{Mendoza_2021}. Experimental evidence based on transport measurements was also reported about this exotic phenomenon~\cite{PhysRevB.94.205418}. The so-called electronic cloaking is owing to the chirality mismatch of states outside and inside a bilayer graphene junction. In fact, electrons impinging onto the potential barrier at normal incidence can tunnel through it as if no localized states were available in their way. Until now there is no evidence of electronic cloaking effect in others 2D materials. 

In this letter, we show that electronic cloaking is also possible in phosphorene. To this end, we study the electronic transport of gated phosphorene junctions along the armchair (AC) and zigzag (ZZ) direction. We find electronic cloaking in the ZZ direction, while in the AC direction the confined states manifest as Fabry-P\'erot resonances in the transmission. We also analyze the hallmarks of the electronic cloaking on the transport properties, including the effects of the bandgap modulation and an applied magnetic field. 

\begin{figure}[htb!]
\includegraphics[width=0.9\textwidth]{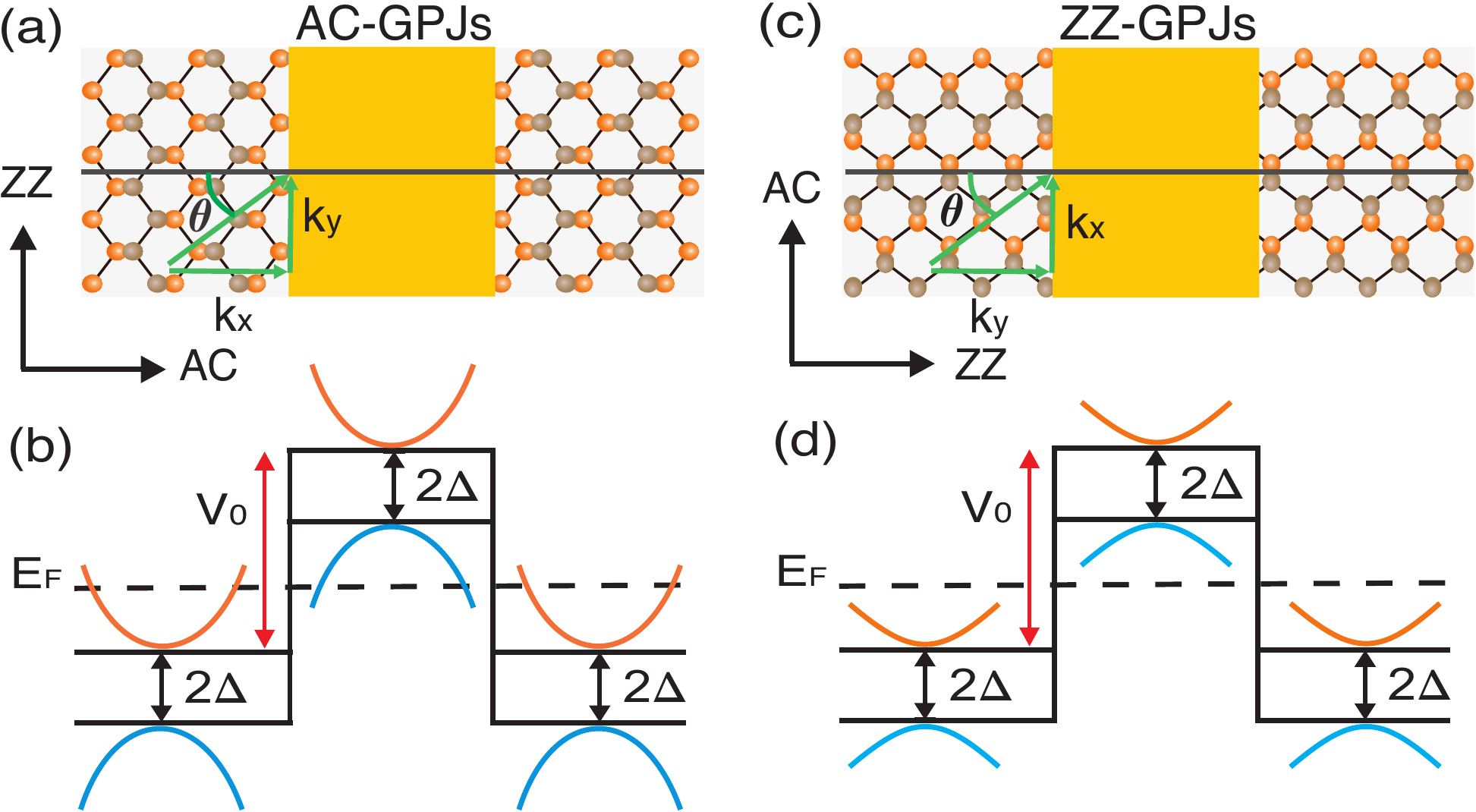}
\caption{\label{Fig1} (a) and (c) Schematic representation of gated phosphorene junctions along the AC and ZZ directions. (b) and (d) Band-edge profile along AC and ZZ directions.} 
\end{figure}

Phosphorene is a 2D semiconductor composed of two atomic layers arranged in a puckered honeycomb lattice. Its experimental fabrication was realized in 2014 by the mechanical exfoliation technique~\cite{Castellanos_Gomez_2014,Carvalho2016}. The phosphorene band structure is anisotropic, with the AC and ZZ as fundamental directions, and a direct bandgap at the $\Gamma$ point~\cite{Ezawa_2014,KouLiangzhi2015}. These characteristics rise the prospects of phosphorene from both the fundamental and technological standpoint. In fact, several exotic phenomena~\cite{Hedayati_Kh_2018,Betancur-Ocampo2019} and possible applications~\cite{Carvalho2016,KouLiangzhi2015} have been reported. Regarding gated phosporene junctions (GPJs), they can be fabricated by placing electrostatic gates over the phosphorene layer as shown in Fig.~\ref{Fig1}(a),(c)~\cite{PhysRevLett.114.066803,Avsar2015ACSNANO}. In the case of AC-GPJs, the charge carriers propagate along the $x$-direction and the electrostatic gate is oriented in the transverse direction, vice versa for ZZ-GPJs. The electrostatic gate induces a potential barrier by shifting the electronic structure, Fig.~\ref{Fig1}(b),(d). 

The electronic transport in GPJs can be modeled with the tight-binding low-energy effective Hamiltonian~\cite{Popovic-PhysRevB.92.035135,PhysRevB.94.165433,De_Sarkar_2017,Betancur-Ocampo2019,Ezawa_2014} 
\begin{equation}\label{E1}
H=\begin{pmatrix}
V({\bf x}) & g^{*}(k) \\
g(k) & V({\bf x})
\end{pmatrix},
\end{equation}

\noindent{where $g(k)=\Delta +\frac{p_{\rm x}^{2}}{2m_{\rm x}}+\frac{p_{\rm y}^{2}}{2m_{\rm y}}+ivp_{\rm x}$, $m_{\rm x}=2\hbar^{2}/(-t_{1}a^{2}+2\delta(2t_{1}a-\delta \Delta))$, $m_{\rm y}=-2\hbar^{2}/(t_{1}b^{2})$, $v=(t_{1}a-\delta \Delta)/\hbar$, $t_{1}=-1.22$eV and $t_{2}=3.665$eV are the hopping energies between the first and second nearest neighbors~\cite{Katsnelson-PhysRevB.89.201408}, $2\Delta=4t_{1}+2t_{2}=2.45$eV is the direct band gap, $a=4.42{\rm \AA}$ and $b=3.27{\rm \AA}$ are the distance of the orthogonal phosphorene basis, $\delta=0.8{\rm \AA}$ is the distance between the two atoms in the unit cell, and $V({\bf x})$ is the potential barrier that can correspond to AC-GPJs $V(x)=V_0$ or ZZ-GPJs $V(y)=V_0$, $V_0$ being the strength of the electrostatic potential. The dispersion relation in the potential barrier region is given by}
\begin{equation}\label{E2}
E=V_{0}\pm \sqrt{\left(\Delta+\frac{\hbar^{2}q_{\rm x}^2}{2m_{\rm x}}+\frac{\hbar^{2}q_{\rm y}^2}{2m_{\rm y}}\right)^2+v^{2}\hbar^{2}q_{\rm x}^2}.
\end{equation}

The transmission is obtained with the help of the numerical stable hybrid matrix method~\cite{BRIONESTORRES2016186}. This method consists in rewrite the eigenvalue problem $H\psi=E\psi$ as Sturm-Liouville matrix equation~\cite{RPA}, ${\bf L}(x)\cdot {\bf F}(x)\equiv d{\bf A}(x)/dx+{\bf Y}\cdot d{\bf F}(x)/dx+ {\bf W}(x)\cdot{\bf F}(x)$, where ${\bf F}(x)$ is the wave function or field, ${\bf A}(x)={\bf B}(x)\cdot d{\bf F}(x)/dx+{\bf P}(x)\cdot{\bf F}(x)$ is the linear differential form, and ${\bf Y}(x)$, ${\bf W}(x)$, ${\bf B}(x)$, ${\bf P}(x)$ are coefficient matrices~\cite{RPA}. This equation and the mathematical process that we will outlined correspond to AC-GPJs. However, a similar equation and process apply for ZZ-GPJs. As we are dealing with a constant potential the solution to the Sturm-Liouville matrix equation can be obtained straightforwardly. The general solution is given by ${\bf F}(x)=\sum_{1}^{4} a_{l}{\bf F}_{l}(x)$ and ${\bf A}(x)=\sum_{1}^{4} a_{l}{\bf A}_{l}(x)$, where ${\bf F}_{l}(x)={\bf F}_{l0}e^{iq_{l}x}$ and ${\bf A}_{l}(x)={\bf A}_{l0}e^{iq_{l}x}$ with amplitudes ${\bf F}_{l0}=(\phi_{l0},\varphi_{l0})^{T}=(g_{l}^{*}(k),E-V_{0})^{T}$ and ${\bf A}_{l0}=(\alpha_{l0},\beta_{l0})^{T}=(-(v\hbar/2+iq_{l}\hbar^{2}/2m_{\rm x})(E-V_{0}),(v\hbar/2-iq_{l}\hbar^{2}/2m_{\rm x})\phi_{l0})^{T}$. The eigenvalues $q_{l} \,(l=1,2,3,4)$ are given by $q_{1}=\sqrt{-\xi_{1}+\xi_{2}}$, $q_{2}=-q_{1}$, $q_{3}=i\sqrt{\xi_{1}+\xi_{2}}$, $q_{4}=-q_{3}$, with $\xi_{1}=2m_{\rm x}^{2}v^{2}/\hbar^{2}+(2m_{\rm x}/\hbar^{2})(\Delta+\hbar^{2}q_{\rm y}^{2}/2m_{\rm y})$, $\xi_{2}=(2m_{\rm x}/\hbar^{2})\sqrt{(E-V_{0})^{2}+m_{\rm x}^{2}v^{4}+2m_{\rm x}v^{2}(\Delta+\hbar^{2}q_{\rm y}^{2}/2m_{\rm y})}$. As we can notice, there are two propagating waves and two evanescent-divergent waves. The solutions in the left and right semi-infinite regions can be obtained readily by taking $V_0=0$, $q_{l}\to k_{l}$, and $q_{\rm y}\to k_{\rm y}$. With the solutions in the different regions of our system we can define the hybrid matrix as \cite{BRIONESTORRES2016186,RPA}:
\begin{equation}\label{E3}
\begin{bmatrix}
           {\bf F}_{\rm L}(x_{ l}) \\
           {\bf A}_{\rm R}(x_{ r}) 
         \end{bmatrix} ={\bf H}(x_{ l},x_{ r})\cdot \begin{bmatrix}
           {\bf A}_{\rm L}(x_{ l}) \\
           {\bf F}_{\rm R}(x_{ r})  
         \end{bmatrix},
\end{equation}

\noindent{where ${\bf H}(x_{ l},x_{ r})$ is the hybrid matrix of the phosphorene junction evaluated at the boundaries of the electrostatic barrier. The field and linear differential form of the left-right semi-infinite region ${\bf F}_{\rm L,R}(x_{ l,r})$ and ${\bf A}_{\rm L,R}(x_{ l,r})$ are also evaluated at the boundaries $x_{l,r}$. Eq. (\ref{E3}) can be written in terms of the transmission and reflection amplitudes as follows $({\rm L}_{2}/{\rm L}_{1},{\rm L}_{4}/{\rm L}_{1}, {\rm R}_{1}/{\rm L}_{1}, {\rm R}_{3}/{\rm L}_{1})^{T} =[{\bf M}_{1}-{\bf H}(x_{ l},x_{ r})\cdot {\bf M}_{2}]^{-1}[{\bf H}(x_{ l},x_{ r})\cdot {\bf M}_{3}-{\bf M}_{4}]$, where ${\bf M}_{1}$, ${\bf M}_{2}$, ${\bf M}_{3}$ and ${\bf M}_{4}$ are auxiliary matrices~\cite{BRIONESTORRES2016186,RPA}. Finally, the transmittance can be calculated as $\mathbb{T}=\left| {\rm R}_{1}/{\rm L}_{1}\right|^{2}$.}

The conductance is evaluated using the relation
\begin{equation}
\mathbb{G}(E)=G_{0}\int_{q^{min}}^{q^{max}}\mathbb{T}(E,q)dq,
\end{equation}

\noindent{where $q$ is the transverse wave vector component, $G_{0}=Le^{2}/\pi h$ is the fundamental conductance factor, with $L$ the width of the system in the transverse direction. The integral runs over the interval $q^{min}<q<q^{max}$. In the case of AC-GPJs $q=k_y$ and $L=L_y$, while for ZZ-GPJs $q=k_x$ and $L=L_x$.}
\begin{figure}[htb!]
\centering
\includegraphics[width=0.9\textwidth]{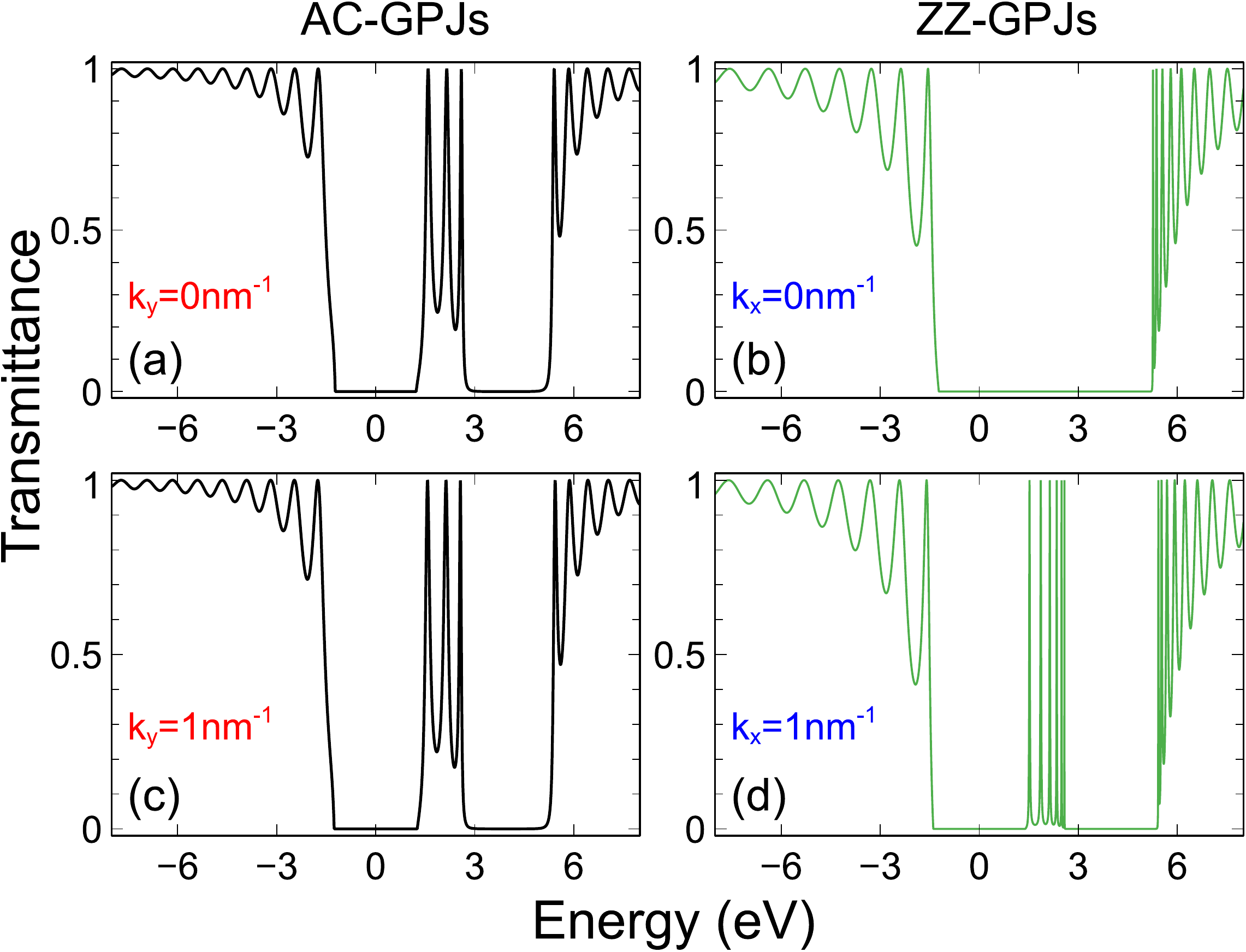}
\caption{\label{Fig2} Transmission as a function of the energy for (first column) AC- and (second column) ZZ-GPJs at (first row) normal and (second row) oblique incidence. The width and strength of the potential barrier in both cases is $d_{B} = 3$nm and $V_0 = 4$eV, respectively.} 
\end{figure}

\begin{figure}[htb!]
\centering
\includegraphics[width=0.9\textwidth]{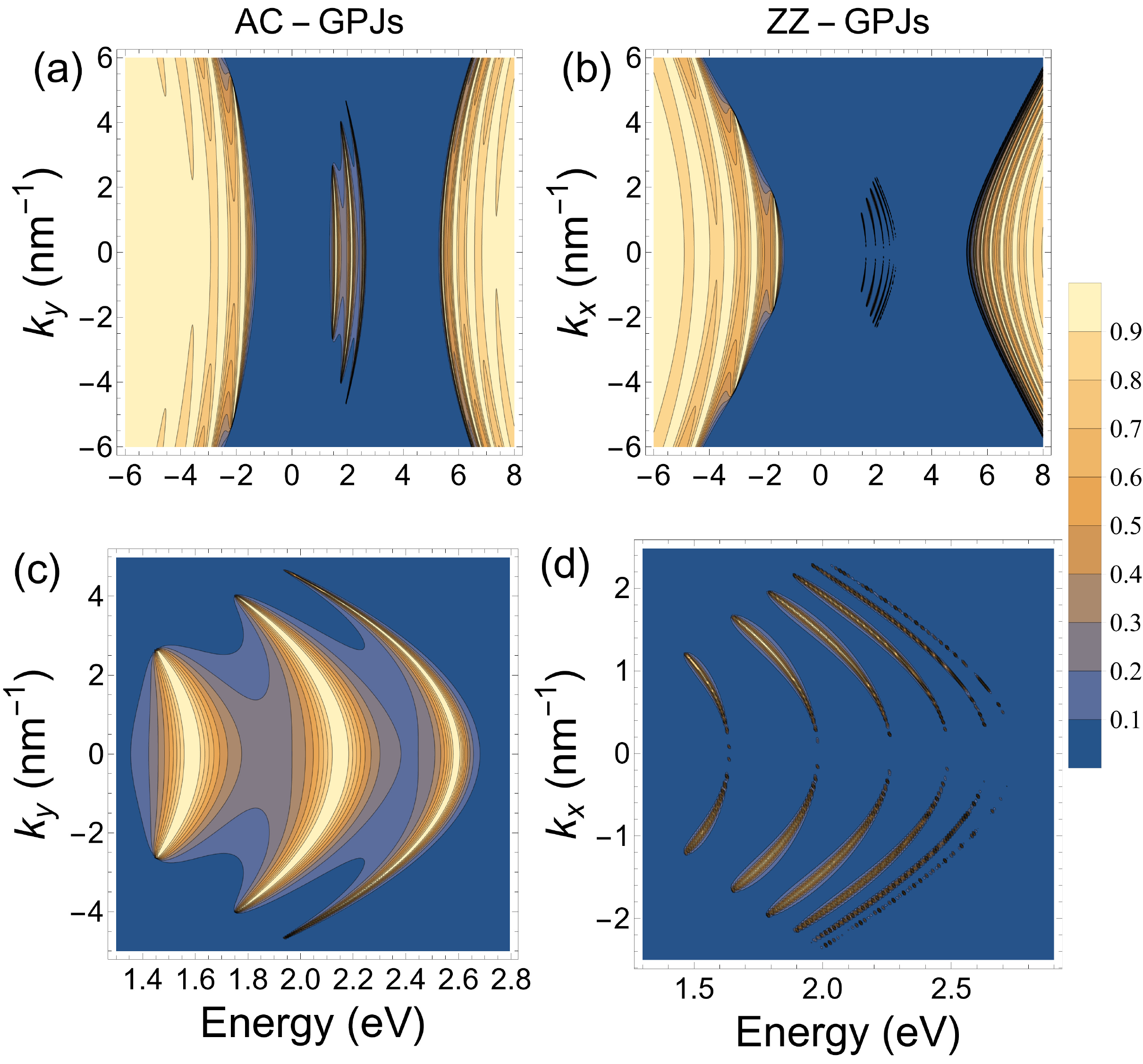}
\caption{\label{Fig3} Transmission maps as a function of the energy and the transverse wave vector for (a) AC- and (b) ZZ-GPJs. (c) and (d) correspond to zooms of (a) and (b), respectively. The structural parameters of the gated junctions are the same as in Fig.~\ref{Fig2}.} 
\end{figure}

\begin{figure}[htb!]
\centering
\includegraphics[width=0.9\textwidth]{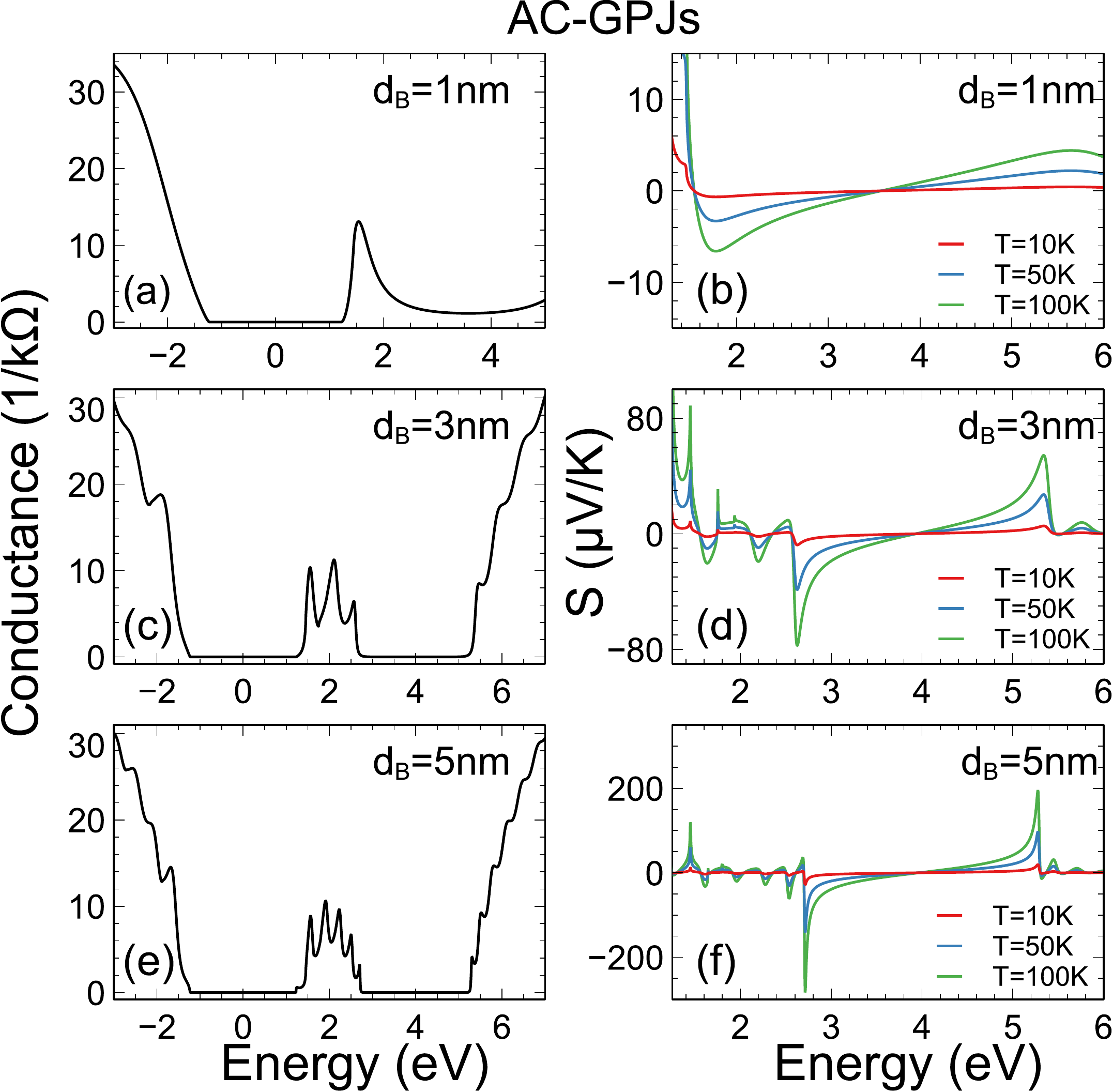}
\caption{\label{Fig4} Conductance (first column) and Seebeck coefficient (second column) as a function of the Fermi energy for AC-GPJs for different barrier widths as indicated. The strength of the electrostatic potential is $V_0 = 4$eV.} 
\end{figure}

As we can notice in Fig.~\ref{Fig1}(c),(d) by considering an appropriate electrostatic potential we can bring hole states inside the barrier of AC- and ZZ-GPJs. In principle, these states will participate in the electronic transport with direct hallmarks in the transmission and conductance. In fact, in the case of AC-GPJs there are transmission resonances at normal ($k_{\rm y}=0$)  and oblique ($k_{\rm y} \neq 0$) incidence, see Fig.~\ref{Fig2}(a),(c). The transmissions maps also corroborate plenty of these resonances in the energy range $1.3{\rm \,eV}<E<2.8{\rm \,eV}$ arranged in semi-circular fashion, see Fig.~\ref{Fig3}(a),(c). The number of resonances at normal incidence coincides with the number of semi-circular regions in the transmission maps. If we fixed $V_0$ and change $d_B$ the number of resonances/semi-circular transmission regions changes as well. For instance, in the case of $V_0=4$eV and $d_B = 1$, 3 and 5nm the number of resonances/semi-circular regions is 1, 3 and 5 respectively, see Fig. S1 in the supplementary information. The conductance also manifests hallmarks associated to the hole states inside the barrier. In particular, the conductance presents sharp peaks located close to the energies of the transmission resonances at normal incidence and that in number correspond to the number of resonances/semi-circular transmission regions, see Fig.~\ref{Fig4}(a),(c),(e). Another quantity that also shows hallmarks of the hole states inside the barrier and could be helpful experimentally is the Seebeck coefficient $S=\frac{\pi^{2}k_{\rm B}^{2}T}{3e}\left.\frac{\partial \ln \mathbb{G}(E)}{\partial E}\right|_{E_{\rm F}}$, where $k_{B}$ is the Boltzmann constant, and $T$ is the average temperature in the thermoelectric device~\cite{SDATTA}. As we can see in Fig.~\ref{Fig4}(b),(d),(f), the Seebeck coefficient shows oscillations in the energy range $1{\rm \,eV}<E<3{\rm \,eV}$, which correspond with the resonances presented in the transmission at normal incidence and with the sharp peaks in the conductance. Similar results are obtained if $d_B$ is fixed and $V_0$ is varied, see Fig. S2.

In the case of the ZZ direction, the electronic transport has been overestimated due to the charge carriers in this direction have been regarded as Schr\"{o}dinger particles~\cite{De_Sarkar_2017,Betancur-Ocampo2019}.  However,  Jung et al.~\cite{JungSung-NM2020} have demonstrated that in reality the charge carriers in phosphorene possess a special pseudospin texture. The pseudospin is a characteristic of two-level quantum systems such as honeycomb lattices~\cite{Montambaux_2009}. The pseudospin is involved in most of the exotic properties of 2D materials, including the electronic cloaking in bilayer graphene \cite{NanGu-PhysRevLett.107.156603,PhysRevB.94.205418}.

The electronic cloaking can be understood with the help of the eigenbasis of $\sigma_{\rm x}$, denoted by $\varphi_{+}$ and $\varphi_{-}$. In this basis the eigenvalue equation $H\psi=E\psi$ results in a system of coupled equations. The coupling between $\varphi_{+}$ and $\varphi_{-}$ is suppressed at normal incidence ($k_{\rm x}=0$), 
\begin{equation}
\left[\frac{d^{2}}{dy^{2}}-\frac{2m_{\rm y}}{\hbar^{2}}(\mp E \pm V_{0}+\Delta)\right]\varphi_{\pm}(y)=0.
\end{equation}  

\noindent $\varphi_{+}$ states represent states that transmit through the potential barrier $V_{0}$, while states $\varphi_{-}$ represent confined states inside the potential well $-V_{0}$. As a result of the decoupling between $\varphi_{+}$ and $\varphi_{-}$ at normal incidence, $\varphi_{+}$ states tunnel through the barrier without interaction with the confined states $\varphi_{-}$. The barrier acts as a {\it cloak} for confined states blocking the transmission completely, i.e., electrons tunnel the barrier as if no localized states were available in their way. In the case of oblique incidence ($k_{\rm x}\neq 0$), $\varphi_{+}$ and $\varphi_{-}$ are coupled, the resultant coupled equations conduce to a differential equation of four order. In this case, in the barrier region we have two propagating waves and two evanescent-divergent waves.

\begin{figure}[htb!]
\centering
\includegraphics[width=0.9\textwidth]{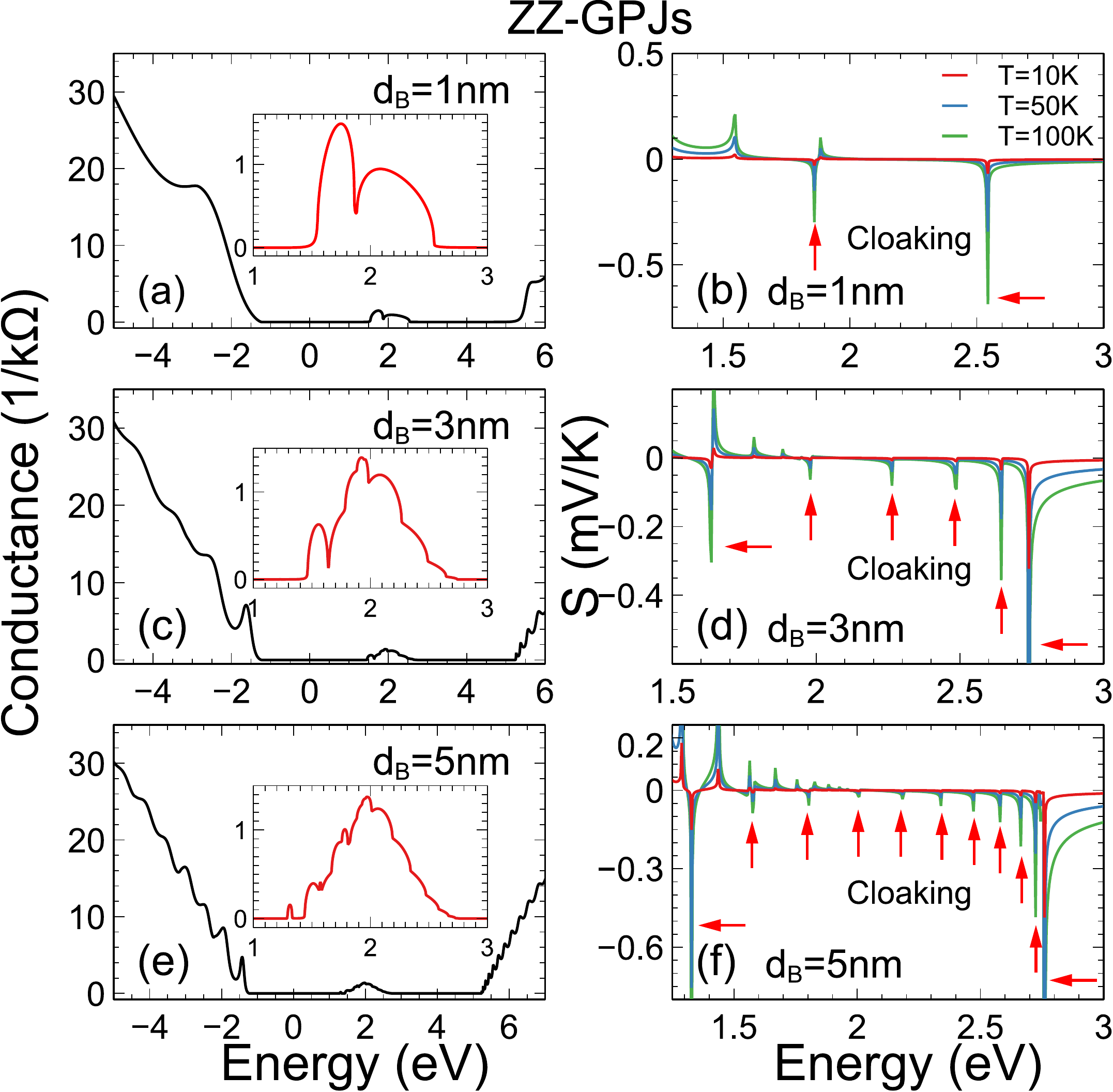}
\caption{\label{Fig5} Conductance (first column) and Seebeck coefficient (second column) as a function of the Fermi energy for ZZ-GPJs at different barrier width as indicated. The strength of the electrostatic potential is $V_0 = 4$eV.} 
\end{figure}

Our numerical calculations confirm the electronic cloaking of confined states in ZZ-GPJs. The transmission at normal incidence ($k_{\rm x}=0$) is totally suppressed in the energy region $1.3{\rm \,eV}<E<2.8{\rm \,eV}$ despite there are hole states inside the barrier, see Fig.~\ref{Fig2}(b). At oblique incidence ($k_{\rm x} \neq 0$), the transmission presents six resonances in the mentioned energy region, see Fig.~\ref{Fig2}(d). We can also appreciate the electronic cloaking in the transmission maps. In Fig.~\ref{Fig3}(b),(d) we can see transmission resonances arranged in semi-circular fashion as in the case of AC-GPJs, however the number of semi-circular regions is greater for ZZ-GPJs. The number of semi-circular regions correspond to the number of resonances in the transmission at oblique incidence. More importantly, the transmission is totally suppressed at normal and nearly normal incidence, manifesting the cloaking effect of the confined states. The transport properties also manifest hallmarks of the electronic cloaking. As we can see in the inset of Fig.~\ref{Fig5}(a),(c),(e), the conductance shows several singular peaks, increasing in number as $d_B$ increases. Moreover, the energy position of the singular peaks coincides with the energy location of the invisible confined sates, constituting a direct hallmark of electronic cloaking. As we mentioned earlier an alternative quantity that can help to confirm the cloaking effect is the Seebeck coefficient. The electronic cloaking results in inverted (negative) spikes in the Seebeck coefficient as characteristic hallmark. We have identified them with red arrows in Fig.~\ref{Fig5}(b),(d),(f). As we increase the barrier width the number of inverted spikes in the Seebeck coefficient increases as well, corresponding to the number of singular peaks in the conductance. Similar results are obtained if we fixed the barrier width and increase systematically the potential, see Fig. S3 in the supplementary information. 

The modulation of the bandgap $2\Delta$ can give us information about the prevalence of the cloaking effect, see Figs. S4, S5, and S6. In particular, it is instructive to know the modifications caused by closing the bandgap $\Delta=0$. In the AC direction, the transmission at normal incidence is oscillatory, not a perfect transmission (Klein tunneling) as in graphene. In the ZZ direction, the cloaking effect is preserved, that is, the transmission remains blocked at normal incidence. In addition, the number of invisible states increases. The conductance rises as the bandgap is reduced, and the sharp and singular peaks of AC- and ZZ-GPJs respectively smear as the bandgap closes. 

We can also use an external magnetic field to see the prevalence of the cloaking effect and identifiable changes on the hallmarks of the electronic cloaking. An external magnetic field can be incorporated via the vector potential ${\bf A}$ (${\bf B}=\nabla \times{\bf A}$) by modifying the momentum vector ${\bf p}\to {\bf p}+e{\bf A}$. We have considered a piecewise vector potential (deltaic magnetic field) acting on the same region as the electrostatic potential. For AC-GPJs the vector potential is ${\bf A}(x ) = A_{y} \hat{y}=(0,B(B_{0}) l_B,0)$, while for the ZZ-GPJs ${\bf A}(y) = A_{x} \hat{x}=(B(B_{0}) l_B,0,0)$. Here, $l_B=\sqrt{\hbar/eB_0}$ is the magnetic length, and $B_0$ the strength of the reference magnetic field. In the case of ZZ-GPJs, the presence of the magnetic field suppresses the cloaking effect, shifting the semi-circular transmission regions until a transmission gap takes place at high magnetic fields, see Fig. S8. The applied magnetic field also transform the singular peaks of the conductance in sharp peaks, shifting them to higher energies and disappearing systematically with the magnetic field until a conductance gap is created at high magnetic fields, see Fig. S9. Similar results are obtained if in addition to the magnetic field the bandgap is closed, see Fig. S10. However, in this case we can choose wider barriers, smaller electrostatic potentials, and more importantly reasonable magnetic fields. In the case of AC-GPJs, the applied magnetic field is not as preponderant as for ZZ-GPJs, requiring stronger magnetic fields to see sizable changes in the transmission and transport properties, see Fig. S7 and S9. 

Finally, it is important to mention that the electronic cloaking takes place only when we use the Hamiltonian given by Eq.~(\ref{E1}). As far as we know, there is no evidence of the cloaking effect when alternative Hamiltonians are used. For instance, Biswas et al.~\cite{Biswas2021SR} used a ${\bf k}\cdot{\bf p}$ Hamiltonian, finding Fabry-P\'erot resonances at normal incidence for ZZ-GPJs. In this context, we hope that our findings motivate experimentalists to corroborate electronic cloaking and theoreticians to study the validity of the different phosphorene Hamiltonians.

In conclusion, the highly directional-dependent pseudospin texture of the charge carriers in phosphorene results in electronic cloaking at normal incidence for ZZ-GPJs, while for AC-GPJs confined states manifest as Fabry-P\'erot resonances in the transmission. The invisible confined states manifest themselves as singular peaks in the conductance and as inverted (negative) spikes in the Seebeck coefficient. The electronic cloaking is insensible to the modulation the bandgap and is lost with the application of a magnetic field. Furthermore, the peaks associated to the cloaking effect are smeared with the bandgap modulation and the applied magnetic field. These characteristics can be used as a hallmark of the electronic cloaking in transport, thermoelectric or magnetotransport measurements. 

\begin{acknowledgments}
I.R.-V. acknowledges CONACYT-Mexico for the financial support through grant A1-S-11655. 
\end{acknowledgments}


\bibliography{bibliography}

\end{document}


\maketitle

\renewcommand{\thefigure}{S\arabic{figure}} 
\setcounter{figure}{0}

\section*{S1. Transmission maps, Conductance and Seebeck coefficient  as a function of Fermi energy for AC- and ZZ-GPJs for different electrostatic potential}

\begin{figure}[htb!]
\centering
\includegraphics[width=0.8\textwidth, angle=90]{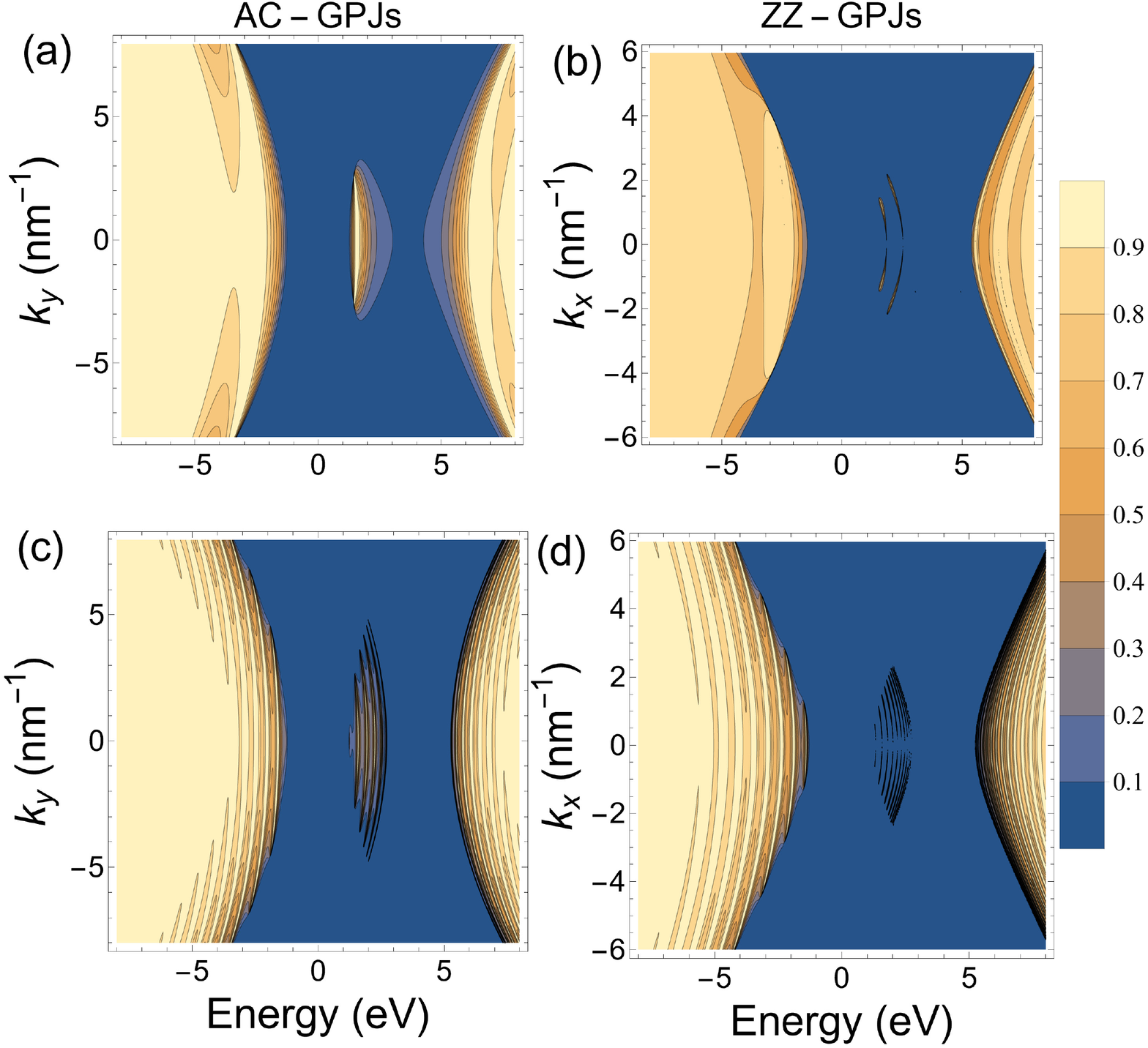}
\caption{\label{Fig1} Transmission maps as a function of the energy for AC- and ZZ-GPJs for different barrier widths as indicated. The strength of the electrostatic potential is $V_{0} = 4$eV.} 
\end{figure}

\begin{figure}[htb!]
\centering
\includegraphics[width=0.9\textwidth]{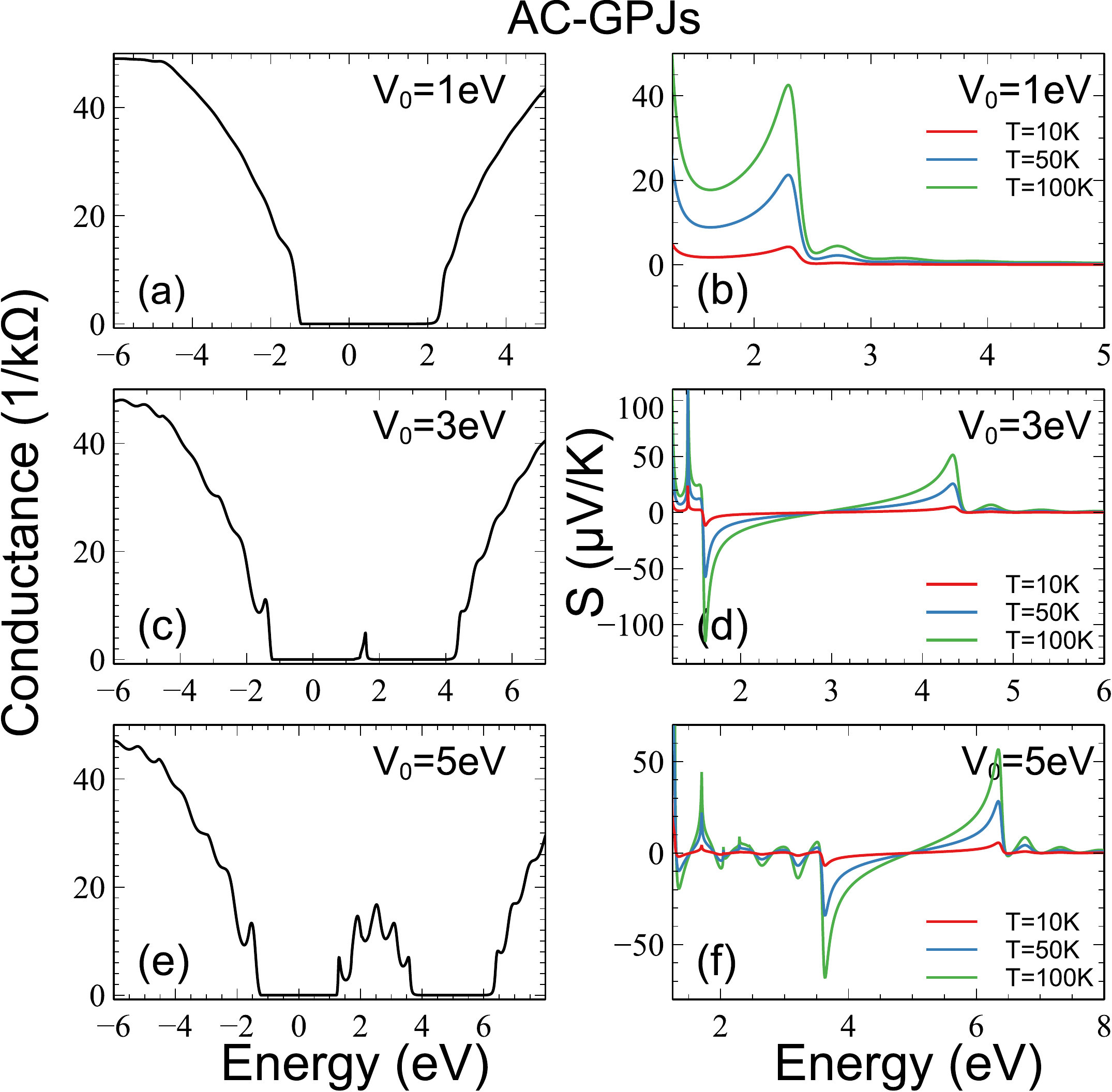}
\caption{\label{Fig2} Conductance (first column) and Seebeck coefficient (second column) as a function of the Fermi energy for AC-GPJs for different electrostatic potential as indicated. The width of the barrier is $d_B = 3$nm. } 
\end{figure}

\begin{figure}[htb!]
\centering
\includegraphics[width=0.9\textwidth]{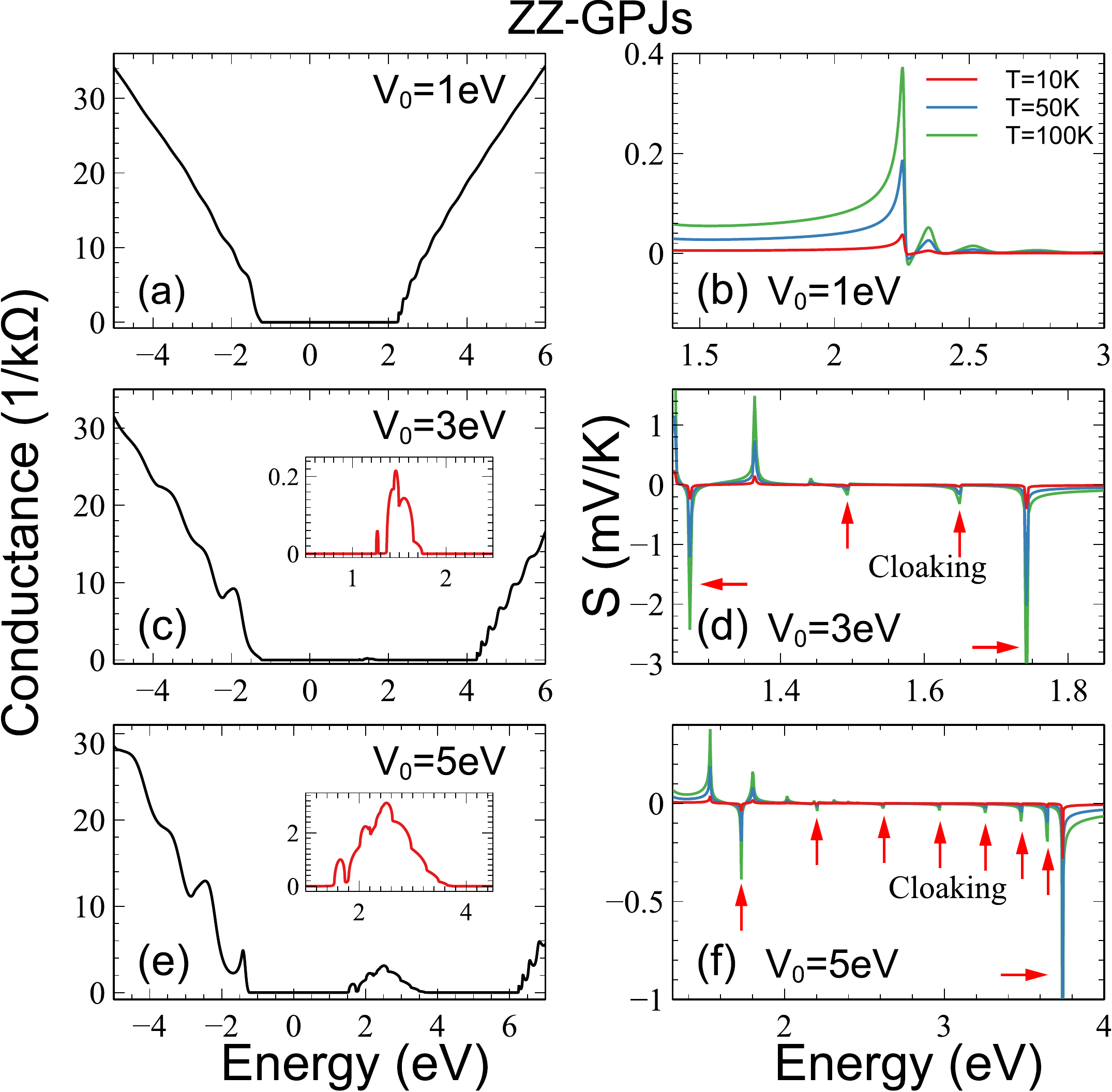}
\caption{\label{Fig3} Conductance (first column) and Seebeck coefficient (second column) as a function of the Fermi energy for ZZ-GPJs for different electrostatic potential as indicated. The width of the barrier is $d_B = 3$nm.} 
\end{figure}

\clearpage
\section*{S2. Transmittance and transmissions maps as a function of the Fermi energy for AC- and ZZ-GPJs for the special case of zero bandgap ($\Delta=0$).}

\begin{figure}[htb!]
\centering
\includegraphics[width=0.9\textwidth]{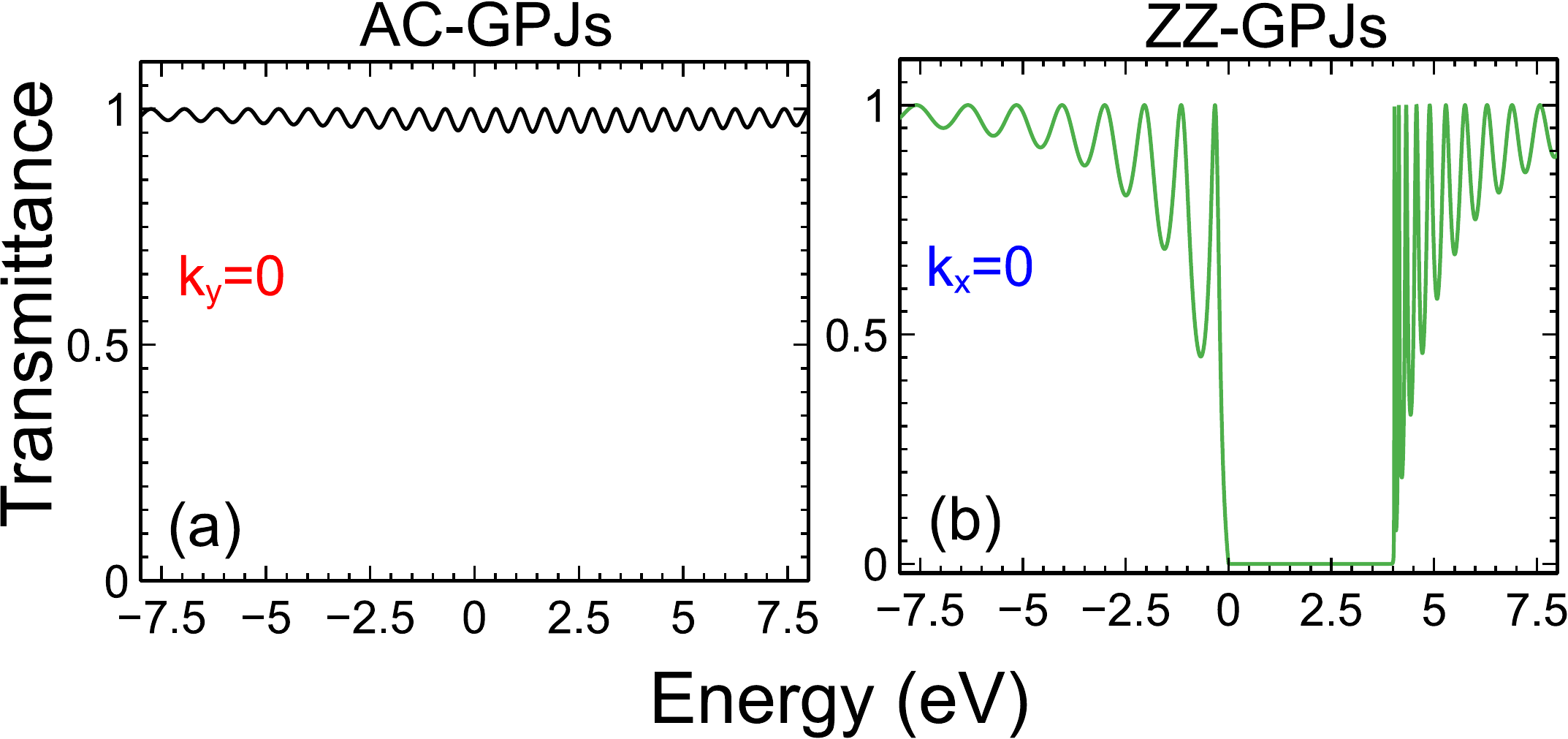}
\caption{\label{Fig4} Transmittance as a function of the Fermi energy for AC (first column) and ZZ (second column) GPJs for zero bandgap ($\Delta=0$) at normal incidence. The width of the barrier is $d_B = 3$nm and the potential is $V_{0} = 4$eV.} 
\end{figure}

\begin{figure}[htb!]
\centering
\includegraphics[width=0.9\textwidth]{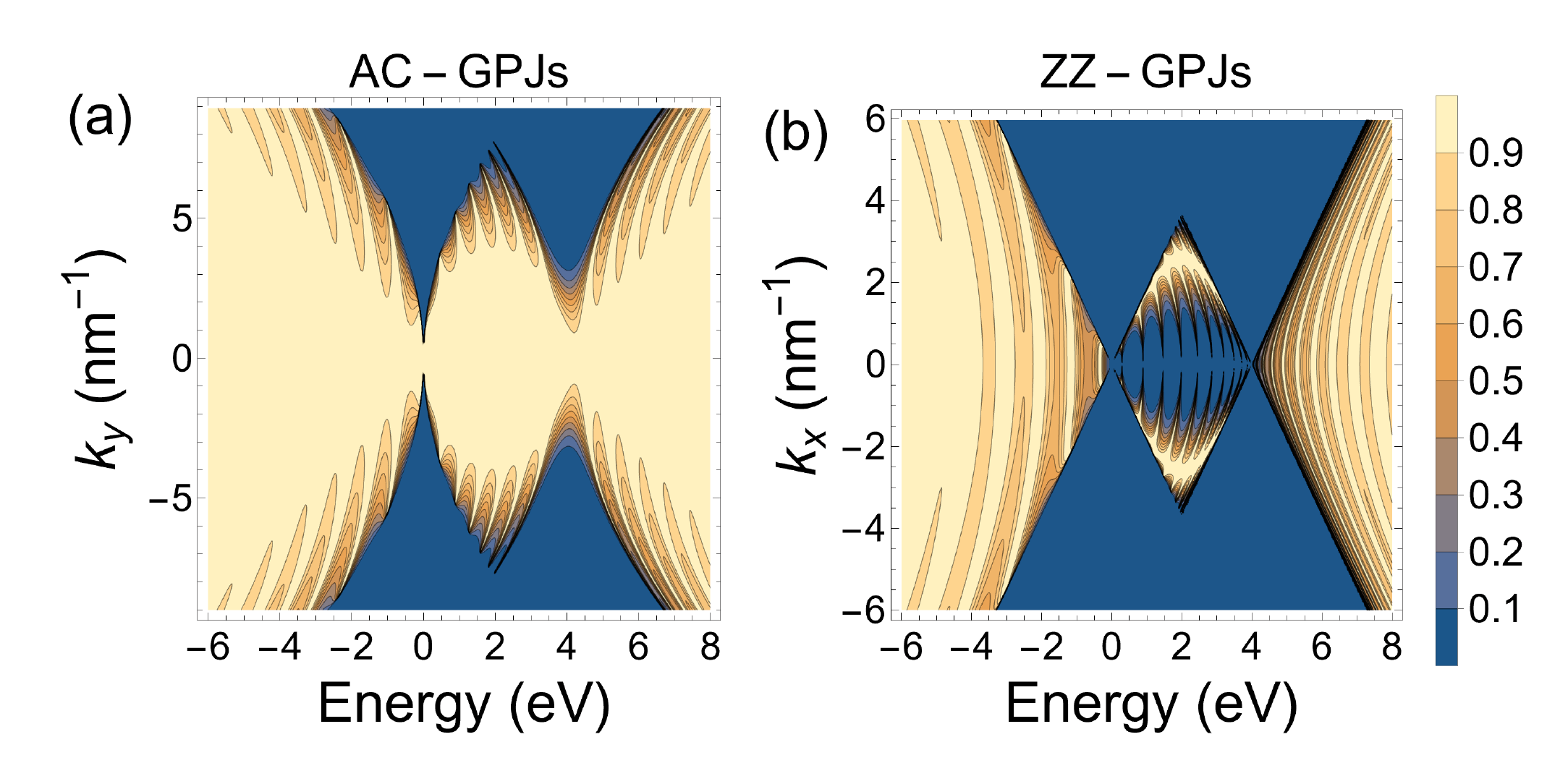}
\caption{\label{Fig5} Transmission maps as a function of the energy for AC- and ZZ-GPJs for the special case of zero bandgap ($\Delta=0$). The width of the barrier is $d_B = 3$nm and the potential is $V_{0} = 4$eV.} 
\end{figure}

\begin{figure}[htb!]
\centering
\includegraphics[width=0.9\textwidth]{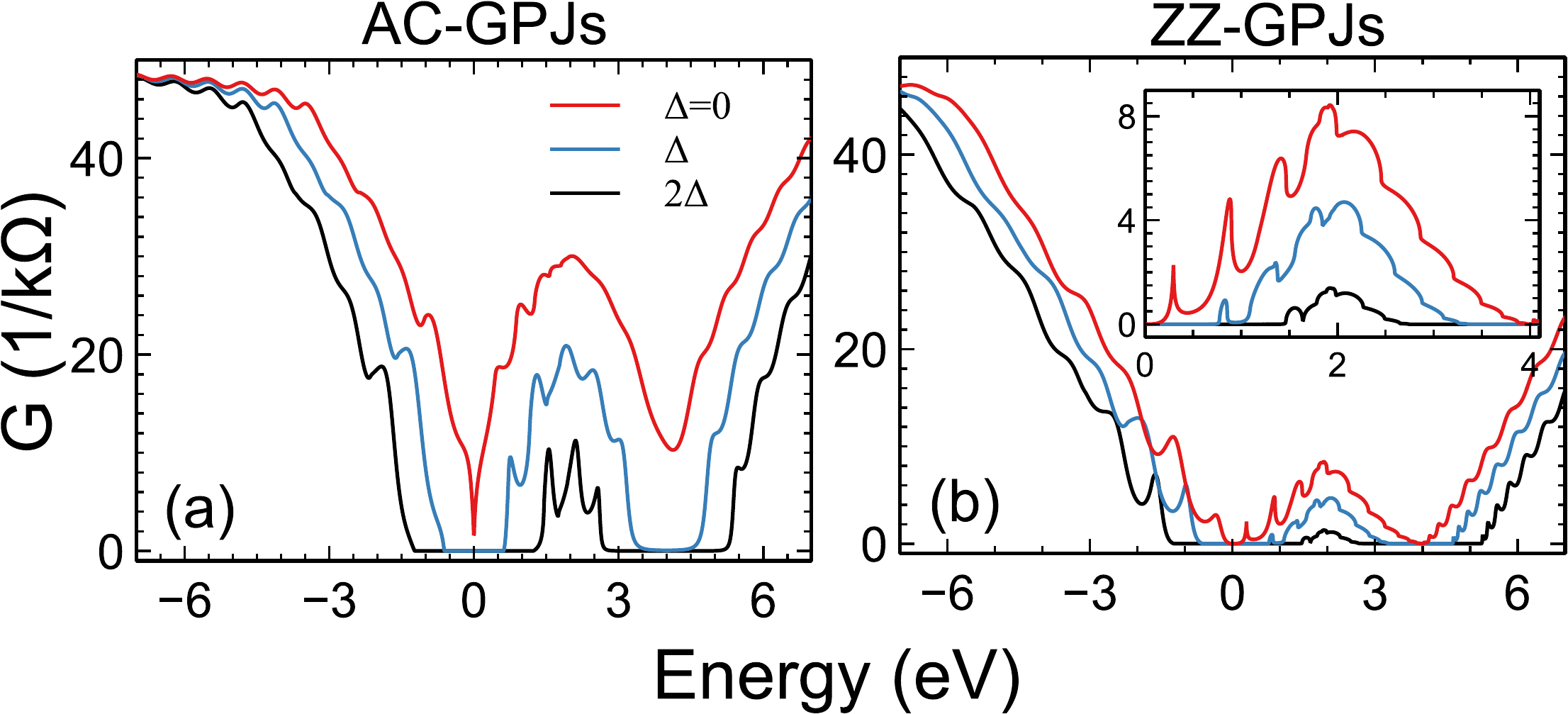}
\caption{\label{Fig6} Conductance as a function of the energy for AC- and ZZ-GPJs for different bandgaps $\Delta$ as indicated. The width of the barrier is $d_B = 3$nm and the potential is $V_{0} = 4$eV.} 
\end{figure}

\clearpage
\section*{S3. Transmission maps as a function of the Fermi energy for AC- and ZZ-GPJs when an external magnetic field is applied.}

\begin{figure}[htb!]
\centering
\includegraphics[width=0.9\textwidth]{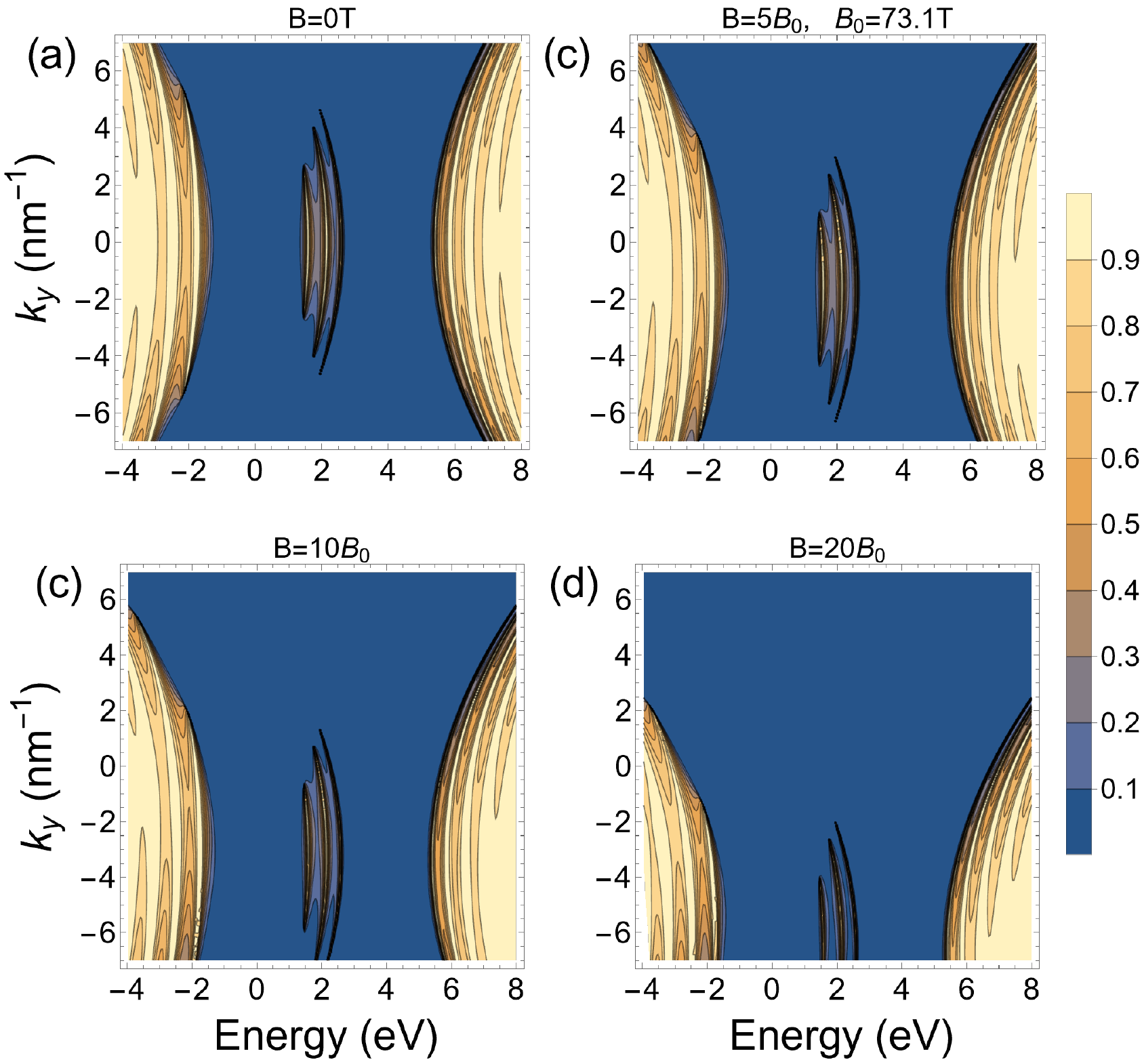}
\caption{\label{Fig7} Transmission maps as a function of the energy for AC-GPJs when an external magnetic field is applied. The width of the barrier is $d_B = 3$nm and the potential is $V_{0} = 4$eV.} 
\end{figure}

\begin{figure}[htb!]
\centering
\includegraphics[width=0.9\textwidth]{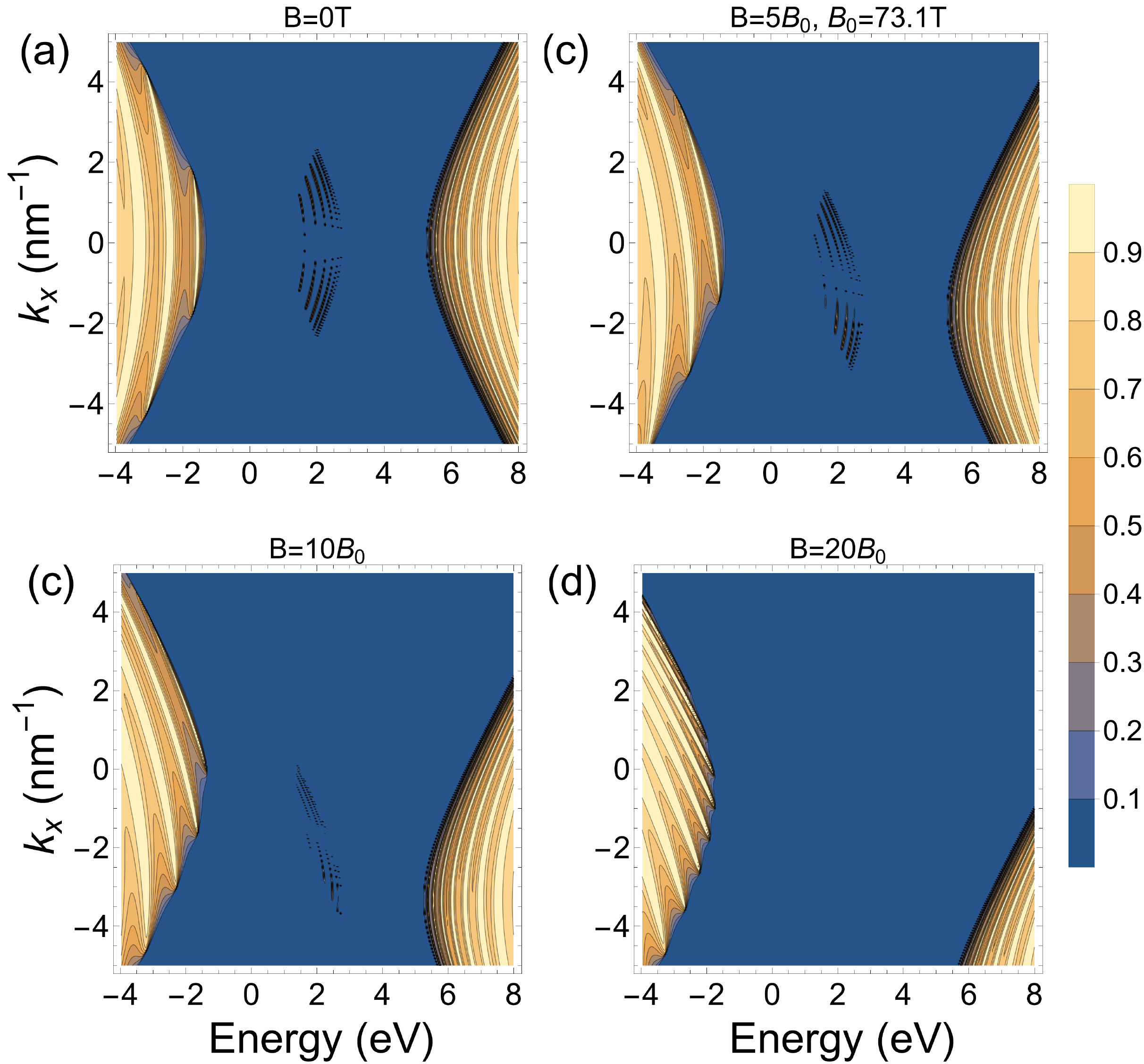}
\caption{\label{Fig8} Transmission maps as a function of the energy for ZZ-GPJs when an external magnetic field is applied. The width of the barrier is $d_B = 3$nm and the potential is $V_{0} = 4$eV.} 
\end{figure}

\begin{figure}[htb!]
\centering
\includegraphics[width=0.9\textwidth]{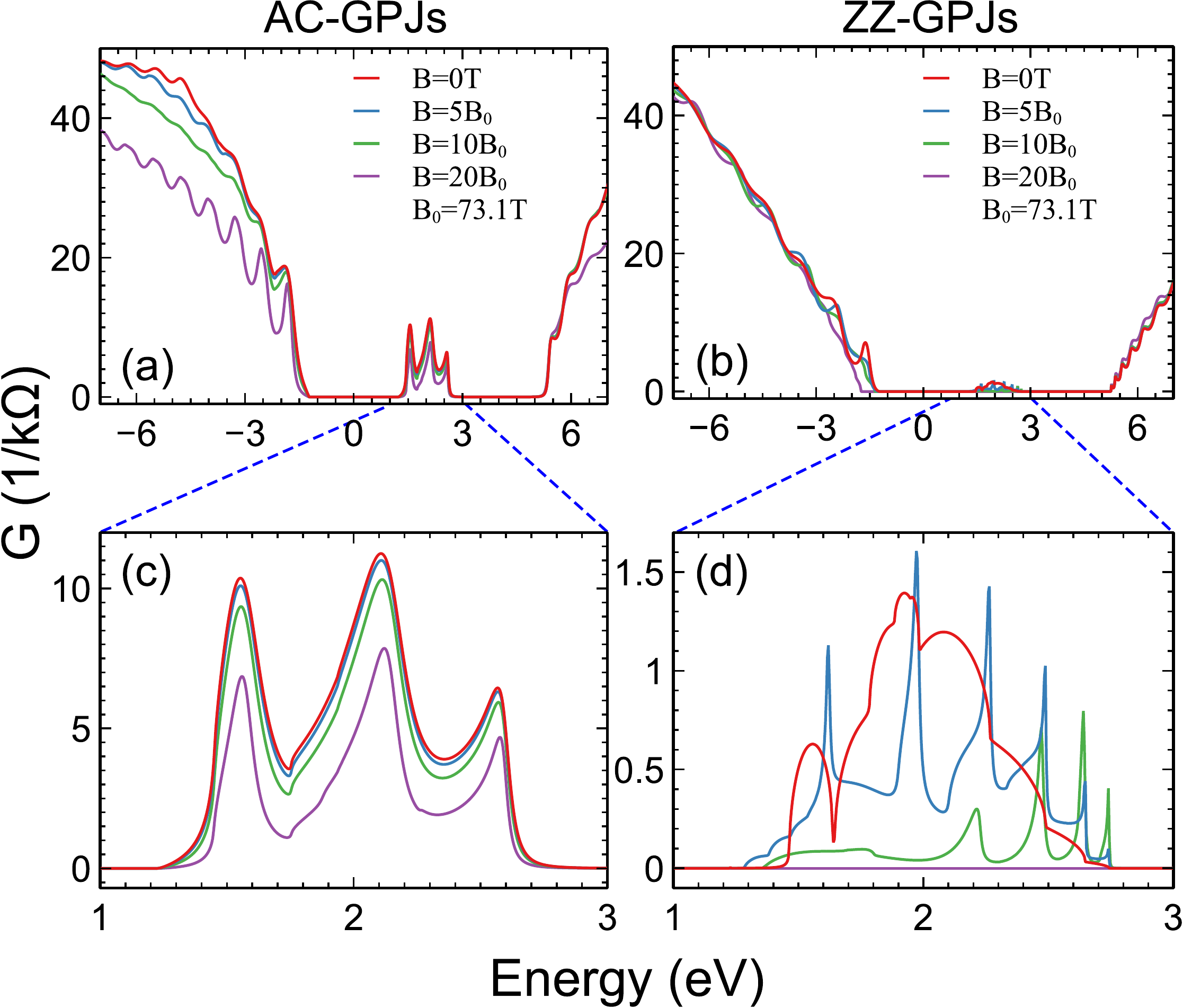}
\caption{\label{Fig9} Conductance as a function of the energy for AC- and ZZ-GPJs when an external magnetic field is applied. The width of the barrier is $d_B = 3$nm and the potential is $V_{0} = 4$eV.} 
\end{figure}

\begin{figure}[htb!]
\centering
\includegraphics[width=0.9\textwidth, angle=90]{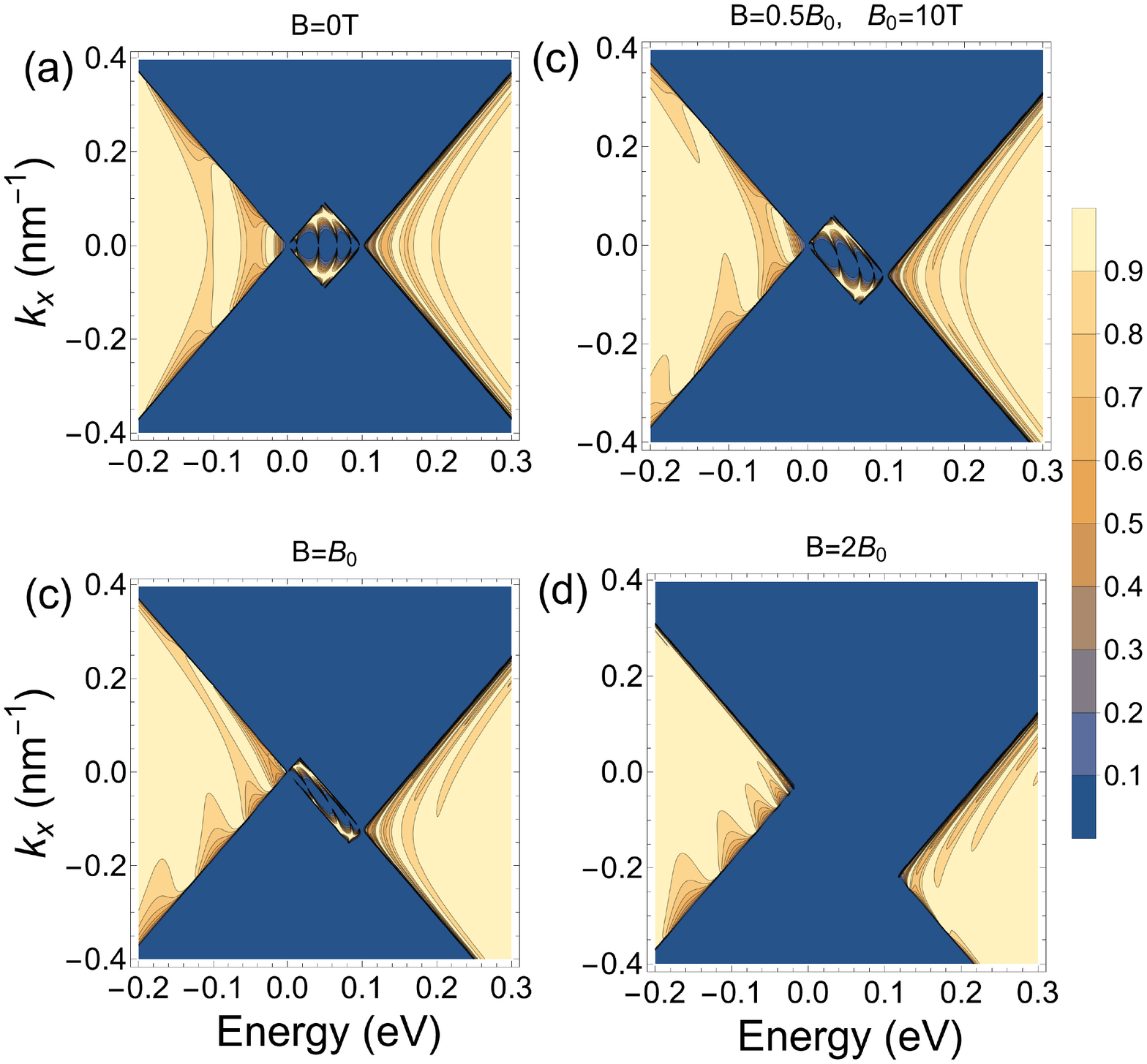}
\caption{\label{Fig10} Transmission maps as a function of the energy for ZZ-GPJs when an external magnetic field is applied. The width of the barrier is $d_B = 8.1$nm, the potential is $V_{0} = 0.1$eV and the bandgap is zero ($\Delta=0$).} 
\end{figure}
